\documentclass[11pt,a4paper]{article}

\usepackage{amssymb,mathtools,amsmath,bm}
\usepackage{graphicx}
\usepackage{hyperref}
\usepackage[utf8]{inputenc}
\usepackage{babel}
\usepackage{caption}
\def\p{{\partial}}
\def\Re{\mathop{\text{Re}}}
\def\Im{\mathop{\text{Im}}}

\newcommand{\Tr}{\mathrm{Tr}}

\title{Multi-point passage probabilities and Green's functions for SLE${}_{8/3}$ }
\author{Oleg Alekseev}
\date{}

\begin{document}
\maketitle

\begin{abstract}We consider a loop representation of the $O(n)$ model at the critical point. When $n=0$ the model represents ensembles of self-avoiding loops (i.e., it corresponds to SLE with $\kappa=8/3$), and can be described by the logarithmic conformal field theory (LCFT) with central charge $c=0$. We focus on the correlation functions in the upper-half plane containing the twist operators in the bulk, and a pair of the boundary one-leg operators. By using a Coulomb gas representation for the correlation functions, we obtain explicit results for probabilities of the SLE${}_{8/3}$ trace to wind in various ways about $N\geq 1$ marked points. When the points collapse pairwise the probabilities reduce to multi-point Green's functions. We propose an explicit representation for the Green's functions in terms of the correlation functions of the bulk 1/3-weight operators, and a pair of the boundary one-leg operators.
\end{abstract}

%%%%%%%%%%%%%%%%%%%%%%%%%%%%%%%%%%

\section{Introduction}
Schramm-Loewner evolution (SLE) provides a conventional framework to study fractal curves or sets growing into simply connected planar domains $\mathcal D\in\mathbb C$~\cite{Sch00}. This approach focuses on constructing measures on random curves that occur in such systems. In the simplest setting of SLE from $x_1$ to $x_2$ (such that $x_1,x_2\in\p \mathcal D$) the measure is generated dynamically by evolving the curve starting from one end point. Conventionally, the domain is taken to be the upper-half plane $\mathbb H=\{z\in\mathbb C:\Im z>0\}$. Then, the curve $\gamma_t$ evolving up to time $t$ (or rather its hull $K_t$) is characterized by the conformal mapping, $g_t:\ \mathbb H\setminus K_t\to\mathbb H$, normalized so that $g_t(z)\sim z+2t/z+O(z^{-2})$ as $z\to\infty$. This function satisfies Loewner equation:
\begin{equation}
\frac{dg_t(z)}{dt}=\frac{2}{g_t(z)-\sqrt\kappa B_t},
\end{equation}
where $B_t$ is a standard Brownian motion, and the real parameter $\kappa$ has a big influence on the geometric properties of SLE${}_\kappa$. The random curves are simple paths provided that $\kappa\leq4$; when $4<\kappa\leq 8$ the curves have no self-intersections but can have double points, while for $\kappa\geq 8$ the curves become space-filling~\cite{RS05}.

Various geometric observables are useful and important in SLE theory. One of the simplest SLE observable is a probability, $P_\kappa(z)$, that the curve passes to the left of a given point $z\in\mathbb H$~\cite{schramm2001}:
\begin{equation}\label{P1-8/3}
P_\kappa(z)=\frac12+\frac{\Gamma(4/\kappa)}{\sqrt\pi \Gamma(\frac{8-\kappa}{2\kappa})}\frac{x}{y}\,{}_2F_1\left(\frac12,\frac{4}{\kappa},\frac32;-\frac{x^2}{y^2}\right),
\end{equation}
where ${}_2F_1(a,b,c;x)$ is a hypergeometric function, and $z=x+i y$. When $\kappa=8/3 $ the probability simplifies to $P_{8/3}(z)=\frac12+\frac{x}{2|z|}$. Later, an analogous formula for the two-point function was predicted by Simmons and Cardy by using conformal field theory (CFT) techniques provided that $\kappa=8/3$~\cite{SC09}. In particular, the probability, $P_{8/3}(z,w)$, that the SLE${}_{8/3}$ curve passes to the left of both points, $z=x+i y$ and $w=u+iv$, has the form:
\begin{equation}\label{P2-8/3}
P_{8/3}(z,w)=P_{8/3}(z)P_{8/3}(w)\left(1+\frac{y}{x+|z|}\frac{v}{u+|w|}G(\sigma)\right),
\end{equation}
with
\begin{equation}
G(\sigma)=1-\sigma\,{}_2F_1\left(1,\frac43;\frac53;1-\sigma\right).
\end{equation}
Here $\sigma=|z-w|^2/|z-\bar w|^2$ is a cross-ration of the points $\{z,w,\bar z,\bar w\}$, and bar stands for complex conjugation.

Let us briefly discuss the result of Simmons and Cardy~\cite{SC09}. Their approach uses an intimate relation between SLE${}_{\kappa}$ and CFT, which allows one to study critical curves using CFT methods~\cite{BPZ}. It is well known that many two-dimensional statistical systems, e.g., $O(n)$ model and percolation, can be mapped to an equivalent loop representation~\cite{Nie84}. Various loop ensembles can be conventionally described in terms of SLE~\cite{Sch00,RS05}. Alternatively, the loop model can be mapped to a height model via Coulomb gas. In the continuum limit the latter model is described by CFT. Hence, it becomes possible to study loop ensembles in the CFT framework. We briefly describe this relation in Section~\ref{s:On}.

An essential part of the Simmons-Cardy construction is the identification of the \textit{twist} operators with the $0$-weight Schramm's operator provided that $\kappa=8/3$. Twist operators at the points $z_i\in\mathbb H$ modify statistical weights of the loops which wind in various ways about these points. Roughly speaking, the correlation function containing a single twist operator at the point $z\in\mathbb H$ counts an expected number of loops which separate $z$ from the boundary. In Section~\ref{s:twist-1} we show, that the correlation function is closely connected to Schramm's formula~\eqref{P1-8/3}. In a similar way, the correlation function containing two twist operators in $\mathbb H$ can be used to count an expected number of loops which separate both point from the boundary. In Ref.~\cite{SC09} Cardy and Simmons showed that this function is closely connected to the two-point probability~\eqref{P2-8/3}.

One purpose of this article is to generalize Simmons-Cardy result to the case of $N\geq3$ points in $\mathbb H$. In this case the system of PDEs which governs the corresponding probabilities is very difficult to solve directly. We take advantage of the CFT technique, namely, the Coulomb gas formalism~\cite{DF84,DF85}, which provides a tractable approach to constructing explicit solutions. As the result, we obtain explicit expressions for probabilities of the SLE${}_{8/3}$ trace to wind in various ways about $N\geq3$ points in $\mathbb H$. Remarkably, this result can be used to study SLE multi-point Green's functions. Indeed, the probability that the SLE trace passes between the points $z_1,z_2\in \mathbb H$ becomes the one-point SLE Green's function as the points collapse to one. Similarly, one expects that the $2N$-point passage probability becomes the $N$-point Green's function as the points $z_1,z_2,\dotsc,z_{2N}$ collapse pairwise.

The structure of the paper is straightforward. In Section~\ref{s:On} we briefly review the $O(n)$ model, which serves as a connection between SLE and CFT. We also introduce the twist and legs operators, and describe their conformal properties. In Section~\ref{s:pp} we use CFT technique to calculating probabilities of SLE${}_{8/3}$ trace to wind in various ways about $1,2,3\dotsc,N$ marked point in the upper-half plane. In particular, in Section~\ref{s:Npoint} we obtain Coulomb gas representation for the $N$-point passage probabilities of SLE curves. Section~\ref{s:Green} is devoted to multi-point Green's functions of SLE${}_{8/3}$ curves in the upper-half plane. We obtain explicit expressions for the Green's functions in terms of the correlation functions of 1/3-weight operators in the bulk, and 1-leg operators on the boundary in $c=0$ logarithmic CFT in $\mathbb H$. Finally, we draw our conclusion.

%%%%%%%%%%%%%%%%%%%%%%%%%%%%%%%%%%

\section{The $O(n)$ model, CFT and SLE}\label{s:On}

%%%%%%%%%%%%%%%%%%%%%%%%%%%%%%%%%%

Let us start with a standard loop representation of the $O(n)$ model with $n$-component spins ${\bm s}(r_i)$, such that ${\bm s}^2(r_i)=1$, on the lattice. The partition function of the $O(n)$ model has the form:
\begin{equation}\label{ZTr-def}
Z=\Tr\prod_{\langle ij\rangle}(1+x {\bm s}(r_i)\cdot {\bm s}(r_j)),
\end{equation}
where $x$ is a parameter of the model, and the product in~\eqref{ZTr-def} is over pairs of nearest neighbors. One can expand the product into a sum of $2^K$ terms (where $K$ is the number of nearest neighbors), so that each term is associated to a graph on the lattice in what follows: the bond between $r_i$ and $r_j$ is included in the graph if the factor $x{\bm s}(r_i)\cdot{\bm s}(r_j)$ appears in the expansion. Note, that only the graphs composed of closed loops contribute to the sum.

The partition function takes a particularly simple form if the model is considered on a honeycomb lattice, where the loops can visit each site a maximum of one time. Because $\Tr\, s_a(r_i)s_b(r_j)=\delta_{ab}$, each loop contributes a total weight $n$ to the partition function. Besides, each occupied bond contributes by the factor $x$. Hence, the partition function is equivalent to
\begin{equation}\label{Z-loops}
Z=\sum_{\Lambda}n^{\mathcal N}x^{\mathcal L},
\end{equation}
where the sum is taken over all closed non-intersecting loop configurations $\Lambda$ on the honeycomb lattice, $\mathcal N$ is the number of loops, and $\mathcal L$ is the total length of loops in each configuration.

The long loops are suppressed for small value of $x$, so that the model flows to vacuum under the renormalization group flow. For large values of $x$ the system flows to a fixed point of densely packed loops. At the boundary between these two regimes there exists a critical point at $x = x_c$ with $x_c = (2+\sqrt{2-n})^{-1/2}$, for which the mean loop length diverges, and the system flows to the dilute fixed point. At this point the model is supposed to be conformally invariant. Hence, it becomes possible to study the $O(n)$ model in the CFT framework.

The loop model can be mapped to the Coulomb gas by replacing a sum over closed loops in~\eqref{Z-loops} by a sum over configurations with oriented loops. This can be achieved by inserting the factors $e^{i\pi\chi}$ ($e^{-i\pi\chi}$) at each vertex where the curve turns to the right (left), and sum over two possible orientations of each loop. As the result, each closed loop on the honeycomb lattice contributes the factor $e^{6\pi i\chi}+e^{-6\pi i\chi}$ to the partition function. Because the contribution of each loop should be $n$, one concludes 
\begin{equation}
n=2\cos 6\pi\chi.
\end{equation}

The model of oriented loops can be mapped further into a height model. The directed loops can be treated as the level lines of a height variable, $h(r)$, on the dual of the lattice, provided that the height variable changes by $\pi$ ($-\pi$) whenever one crosses a loop pointing to the right (left). There exists a one-to-one correspondence between a given configuration of heights and a unique graph of oriented loops. One can argue that under the renormalization group flow the height model flows into a free field theory with the action $S[h(r)]=(g(n)/4\pi)\int(\p h(r))^2d^2r$, where $g(n)$ is a constant determined by $n$. Hence, it becomes possible to use field theoretical methods in order to make precise calculations in the continuum limit of the model.

As explained, the $O(n)$ model describes ensembles of closed paths on the lattice. One can argue, that in the continuum limit the measure on the curves is given by SLE${}_\kappa$, and (see, e.g., Ref~\cite{KN04})
\begin{equation}
n=2\cos\frac{(\kappa-4)\pi}{\kappa},
\end{equation}
where $2<\kappa<4$ for a dilute phase, and $4<\kappa$ for a dense phase. The mentioned correspondence between the $O(n)$ model and CFT implies that the loop models can be described by rational CFTs with central charge and conformal weights given by
\begin{equation}\label{h-rs}
c=\frac{(6-\kappa)(3\kappa-8)}{2\kappa}, \qquad h_{r,s}=\frac{(\kappa r-4s)^2-(\kappa-4)^2}{16\kappa}.
\end{equation}
Below, we consider only the dilute regime with $2<\kappa<4$.

Note, that the case $\kappa=8/3$ corresponds to the logarithmic CFT (LCFT) with the central charge $c=0$. LCFTs are characterized by presence of logarithmic structure in the operator product expansion explained by indecomposable representations that occur in fusion of primary operators~\cite{Gur13,EF06}. In other words, there exist primary operators with degenerate scaling dimension constituting a Jordan block structure.

The so-called twist operators introduced in Ref.~\cite{GC06} play a crucial role in Simmons-Cardy construction of the probabilities for the SLE trace to wind in various ways about marked points in $\mathbb H$. A pair of twist operators changes the weights of all loops that separate them. Because the weights of the loops separating the twist operators is $-n$, the partition function for the loop model in presence of twist operators takes the form:
\begin{equation}
Z=\sum_{\Lambda}(-1)^{\mathcal N_s}n^{\mathcal N} x^{\mathcal L},
\end{equation}
where $\mathcal N_s$ is the number of loops separating the twist operators. Hence, the twist operators can be used to count loops with weights $-n$ rather than $n$. The scaling dimension of twist operators can be calculated explicitly~\cite{GC06}. Remarkably, the twist operators correspond to CFT primary fields\footnote{Here and below, we adapt the notation from Ref.~\cite{SC09}}, $\Phi_{2,1}$, and, therefore, their Ka\v{c} weights read:
\begin{equation}
h_{twist} = h_{2,1} = \frac{3\kappa-8}{16},
\end{equation}
The twist operators are spinless, so that the antiholomorphic dimension coincides with the holomorphic one, $\bar h_{twist}=h_{twist}$.

Another set of operators considered in the SLE/CFT correspondence are the so-called boundary $K$-leg operators anchoring SLE traces to the boundary of the domain. In the Coulomb gas framework these operators change the boundary conditions by $K$ steps within $\epsilon$-neighbor of their insertion, and can be identified with the boundary primary operators, $\Phi_{1,K+1}$, with the weights:
\begin{equation}
h_{K-leg}=h_{1,K+1}=\frac{K(4+2K-\kappa)}{2\kappa}.
\end{equation}

We will use a pair of the $1$-leg boundary operators at the points $x_1,x_2\in\mathbb R$ to encode the SLE process in $\mathbb H$ from $x_1$ to $x_2$. The two-point correlation function of such operators is fixed by scale invariance,
\begin{equation}\label{H0}
H_0(x_1,x_2)=\langle\Phi_{1,2}(x_1)\Phi_{1,2}(x_2)\rangle_{\mathbb H}=(x_2-x_1)^{-2h_{1,2}},
\end{equation}
where $\langle\cdots\rangle_{\mathbb H}$ denotes the correlation function in $\mathbb H$, and we set the normalization constant to $1$ by choosing an appropriate normalization of the fields.

%%%%%%%%%%%%%%%%%%%%%%%%%%%%%%%%%%

\section{The passage probabilities of the SLE${}_{8/3}$ trace}\label{s:pp}

%%%%%%%%%%%%%%%%%%%%%%%%%%%%%%%%%%

\subsection{Anchored correlation functions with a twist operator}\label{s:twist-1}

%%%%%%%%%%%%%%%%%%%%%%%%%%%%%%%%%%

In Ref.~\cite{SC09} it is shown, that the correlation function containing a pair of the 1-leg boundary operators, $\Phi_{1,2}$, and the bulk twist operator, $\Phi_{2,1}$, determines probabilities of the SLE${}_{8/3}$ trace to wind in various ways about the point in $\mathbb H$. This result can be easily generalized to the case of $N\geq2$ points. We start this section by rederiving the famous Schramm's formula for the left/right passage probability of the SLE${}_{8/3}$ trace. It is closely connected to the correlation function of the boundary $1$-leg operators in presence of the twist defect at the point $z\in\mathbb H$~\cite{SC09}:
\begin{equation}\label{H1zz-def}
H_1(z,\bar z;x_1,x_2)=\langle \Phi_{2,1}(z,\bar z)\Phi_{1,2}(x_1)\Phi_{1,2}(x_2)\rangle_{\mathbb H},
\end{equation}
where bar stands for complex conjugation, and $x_1,x_2\in \mathbb R$. As per usual CFT approach, the correlation function $H_1(z,\bar z,x_1,x_2)$ in $\mathbb H$ can be represented as the correlation function in $\mathbb C$~\cite{Cardy84},
\begin{equation}\label{H1*}
H_1(z,z^*;x_1,x_2)=\langle\Phi_{2,1}(z)\Phi_{2,1}(z^*)\Phi_{1,2}(x_1)\Phi_{1,2}(x_2)\rangle,
\end{equation}
subjected to certain constraints on $\mathbb R$ specified below. By $\langle\cdots\rangle$ in \eqref{H1*} we denoted the correlation function in the complex plane $\mathbb C$, and the points $z,z^*$ are treated as the independent variables (one sets $z^*=\bar z$ at the end of the computation).

CFT methods allow one to derive a set of second orders PDEs satisfied by the correlation functions containing null state operators, e.g., $\Phi_{1,2}$ and $\Phi_{2,1}$~\cite{BPZ}. In particular, one can show that the correlation function~\eqref{H1*} satisfies the following equations:
\begin{equation}\label{H1-system}
\begin{aligned}
\left[\frac{3\p_z^2}{2(1+2h_{2,1})}-\frac{h_{2,1}}{(z^*-z)^2}+\frac{\p_{z^*}}{z^*-z}-\frac{h_{1,2}}{(x_1-z)^2}+\right.\\ 
\left.\frac{\p_{x_1}}{x_1-z}-\frac{h_{1,2}}{(x_2-z)^2}+\frac{\p_{x_2}}{x_2-z}\right]H_1 =0,\\
\left[\frac{3\p_{x_1}^2}{2(1+2h_{1,2})}-\frac{h_{2,1}}{(z^*-x_1)^2}+\frac{\p_{z^*}}{z^*-x_1}-\frac{h_{2,1}}{(z-x_1)^2}+\right.\\
\left.\frac{\p_{z}}{z-x_1}-\frac{h_{1,2}}{(x_2-x_1)^2}-\frac{\p_{x_2}}{x_2-x_1}\right]H_1 =0.
\end{aligned}
\end{equation}
These equations have the common solution,
\begin{equation}\label{Pi1-G}
H_{1}(z,z^*;x_1,x_2)=(z-z^*)^{-2h_{2,1}}(x_2-x_1)^{-2h_{1,2}} G_1(\eta),
\end{equation}
where $G_1(\eta)$ is the function of the cross ratio $\eta$:
\begin{equation}\label{G1}
G_1(\eta)=\frac{2-\eta}{2\sqrt{1-\eta}},\qquad \eta=\frac{(z- z^*)(x_2-x_1)}{(z-x_1)( x_2-z^*)}.
\end{equation}
The function $G_1(\eta)$ has a branch cut from $1$ to $\infty$. By noting that $1-\eta=[(z^*-x_1)(x_2-z)]/[(z-x_1)(x_2-z^*)]$ and setting $\{z,z^*,x_1,x_2\}\to\{z,\bar z,0,\infty\}$, we conclude, that the choice of the branch of the square root is determined by the argument of $z$. We obtain
\begin{equation}\label{H1-res}
\frac{H_{1}(z,\bar z;0,\infty)}{H_0(0,\infty)}=(2\Im z)^{1-3\kappa/8}\frac{\Re z}{|z|},
\end{equation}
where $H_0$ is the two-point function of the boundary 1-leg operators~\eqref{H0}.

Remarkably, one can obtain exact solution to the system of equations~\eqref{H1-system} via Coulomb gas formalism introduced by Dotsenko and Fateev~\cite{DF84,DF85}. In this approach one uses a representation of the conformal fields in terms of the vertex operators build from a free boson with specific boundary conditions. In particular, there exists a one-to-one correspondence between the primary fields $\Phi_{r,s}(z)$ with the conformal weights~\eqref{h-rs} and the vertex operators
\begin{equation}\label{V-def}
V_{r,s}(z)\equiv V_{\alpha_{r,s}}=e^{i\sqrt2 \alpha_{r,s} \varphi(z)},
\end{equation}
where $\varphi(z)$ is a free boson specified by the two-point function $\langle\varphi(z)\varphi(w)\rangle=-\ln(z-w)$, and $\alpha_{r,s}$ is the so-called charge of the vertex operator:
\begin{equation}
\alpha_{r,s}=\frac12(1-r)\alpha_-+\frac12(1-s)\alpha_+.
\end{equation}
Here $\alpha_\pm$ are determined by the central charge $c$ of CFT as follows:
\begin{equation}\label{c-alpha}
c=1-24 \alpha_0^2,\qquad \alpha_++\alpha_-=2\alpha_0,\qquad \alpha_+\alpha_-=-1.
\end{equation}
Below, we will use the parametrization relevant for SLE/CFT correspondence:
\begin{equation}\label{alpha_pm}
\alpha_+=\frac{2}{\sqrt\kappa},\qquad \alpha_-=-\frac{\sqrt{\kappa}}{2}.
\end{equation}

We refer to $\alpha_0$ in~\eqref{c-alpha} as to the background charge, because Coulomb gas formalism implies an existence of the chiral operator with the charge $-2\alpha_0$ at infinity. The background charge specifies the conformal dimension of the vertex operator, $V_\alpha(z)$, as follows:
\begin{equation}\label{h-alpha}
h_\alpha=\alpha(\alpha-2\alpha_0).
\end{equation}
Note, that the conformal dimension~\eqref{h-alpha} is invariant under $\alpha\to2\alpha_0-\alpha$, so that the vertex operators $V_\alpha$ and $V_{2\alpha_0-\alpha}$ have the same dimensions. Therefore, the conformal field $\Phi_{r,s}$ can be associated to two different vertex operators, $V_{r,s}$ and $V_{-r,-s}$, implying that the correlation functions of conformal fields may be evaluated in several different but equivalent ways.

Because the two-point function of the free boson $\varphi(z)$ has a simple form, the correlation function of vertex operators can be written as
\begin{equation}\label{VV-correlation} 
\langle V_{\alpha_1}(z_1)V_{\alpha_2}(z_2)\cdots V_{\alpha_n}(z_n)\rangle=\prod_{i<j}(z_i-z_j)^{2\alpha_i \alpha_j},
\end{equation}
provided that the following \textit{neutrality condition} is satisfied (otherwise, the correlation function vanishes):
\begin{equation}\label{neutrality}
\sum_{i=1}^n \alpha_i = 2\alpha_0.
\end{equation}
Hence, any multi-point correlation function of vertex operators is nontrivial if and only if the charges satisfy the neutrality condition.

Let us consider the four-point function of vertex operators supposed to realize the four-point correlation function of primary fields~\eqref{H1*}. Note, that it is impossible to write a product of four vertex operators made of $V_{1,2}$, $V_{2,1}$, $V_{-1,-2}$, and $V_{-2,-1}$ that satisfies the neutrality condition. In order to circumvent the problem one can insert a sufficient number of \textit{screening charges}, $Q^\pm$, into the correlation function,
\begin{equation}\label{Qpm}
Q^\pm = \oint_\gamma V_{\pm}(u)du,
\end{equation}
where $\gamma$ is a contour in the complex plane, and $V_{\pm}=V_{\alpha_{\pm}}$. Because the conformal dimension of $V_{\pm}$ is $h_{\alpha_\pm}=1$, the screening charges $Q^\pm$ have vanishing conformal dimensions. Therefore, these charges are invariant under conformal mappings. Inserting $Q^\pm$ an integer number of times in a correlation function of vertex operators will not affect its conformal properties, but will force the neutrality condition to be satisfied.

In the Coulomb gas formalism, the four-point correlation function~\eqref{H1*} is realized via the correlation function of vertex operators with a single screening charge $Q^-$:
\begin{equation}\label{F1-def}
H_1(z,z^*;x_1,x_2;\gamma)=N_1\langle V_{2,1}(z)V_{2,1}(z^*)V_{1,2}(x_1)V_{-1,-2}(x_2)Q^-\rangle,
\end{equation}
where the normalization constant $N_1$ reads depends on the integration contour in the screening charge~\eqref{Qpm}. Eq.~\eqref{VV-correlation} together with~\eqref{Pi1-G} leads to the following explicit expression
\begin{equation}\label{H1*-int}
H_1(z,z^*;x_1,x_2;\gamma)=\frac{N_1\eta^{2h_{2,1}+\kappa/8}\mathcal J_1(\eta)}{(z - z^*)^{2h_{2,1}}(x_1-x_2)^{2h_{1,2}}(1-\eta)^{1/2}},
\end{equation}
where $h_{r,s}=h_{\alpha_{r,s}}$, the cross-ratio $\eta$ was introduced in~\eqref{G1}, $\mathcal J_1(\eta)$ denotes the integral,
\begin{equation}\label{J1-def}
\mathcal J_1(\eta)=\int_0^\eta u^{-\kappa/4}(1-u)(u-\eta)^{-\kappa/4 } du,
\end{equation}
along simple curve connecting $0$ and $\eta$, and the normalization factor reads:
\begin{equation}\label{N1-def}
N_1=\frac{e^{2i\pi h_{2,1}}\Gamma(2-\kappa/2)}{[\Gamma(1-\kappa/4)]^2}.
\end{equation}

Let us briefly comment on the choice of the integration path in~\eqref{H1*-int}. In order to guarantee that~\eqref{H1*-int} satisfies the system of equations~\eqref{H1-system}, the integration contour must be closed. Besides, it must surround at least one of the branch points of the integrand. If the powers of the branch points are irrational (as is usually the case), then the winding number of the contour around each of the points must be zero in order for it to close. The Pochhammer contour is the simplest such contour. However, it is convenient to replace the Pochhammer contour with the path joining the endpoints according to the rule:
\begin{multline}
\oint_{\gamma(z_1,z_2)}f(z_1,z_2,\dotsc;u)du=\\
=4e^{i\pi(\beta_1-\beta_2)}\sin\pi \beta_1\sin\pi \beta_2\int_{z_1}^{z_2}f(z_1,z_2,\dotsc;u)du,
\end{multline}
where $f(z_1,z_2,\dotsc;u)=\prod(u-z_i)^{\beta_i}$, and the monodromy factors, $\beta_1$ and $\beta_2$, are greater than negative one.

After substituting $u=\eta t$ in the integrand of~\eqref{H1*-int} we find
\begin{equation}\label{F1-int}
\frac{H_1(z,\bar z;x_1,x_2)}{H_0(x_1,x_2)}=(2\Im z)^{1-3\kappa/8}\frac{2-\eta}{2\sqrt{1-\eta}}.
\end{equation}
By taking into account~\eqref{G1}, we conclude that the correlation function obtained in the Coulomb gas formalism coincides with~\eqref{H1-res}.

%%%%%%%%%%%%%%%%%%%%%%%%%%%%%%%%%%

\subsection{The 1-point SLE${}_{8/3}$ passage probabilities}

Below, we consider the correlation function~\eqref{F1-int} provided that $\kappa=8/3$. This case is rather tricky, because it corresponds to $c=0$ LCFT. One can argue that two operators, $\Phi_{1,2}$ and $\Phi_{2,1}$, can not be present in the theory simultaneously because of the following reason: the modules generated by these fields lead to existence of staggered modules with different logarithmic coupling, which contradicts to the conformal invariance~\cite{MR07,MR08}. However, both these operators are contained in the correlation function~\eqref{H1zz-def}. Simmons and Cardy suggested, that both operators can coexist in the theory provided that one logarithmic partner appears in the bulk, while the other one on the boundary only~\cite{SC09}.

When $\kappa = 8/3$ we have $n = 0$, so that the $O(n)$ model describes an ensemble of self-avoiding walks (loops) in $\mathbb H$. At $n=0$ all loops are suppressed, and the partition function~\eqref{Z-loops} reads $Z=1$. In presence of the 1-leg boundary operators at the points $x_1,x_2$ we only have those configurations with a self-avoiding path connecting the boundary points $x_1$ and $x_2$. The total weight of these configurations is given by the correlation function of the 1-leg operators at the boundary:
\begin{equation}\label{H0-weight}
H_0(x_1,x_2)=\langle \Phi_{1,2}(x_1)\Phi_{1,2}(x_2)\rangle_{\mathbb H}=(x_1-x_2)^{-5/4}.
\end{equation}

\begin{figure}[t]
\centering
\includegraphics[width=.6\columnwidth]{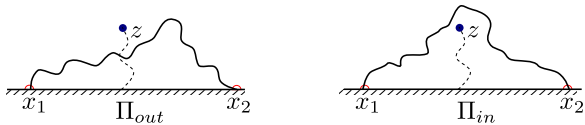}
\caption{\label{P_gamma} The two weights, $\Pi_{out}$ and $\Pi_{in}$, of the upper-half plane SLE${}_{8/3}$ with respect to the twist defect at the point $z$: the weight $\Pi_{out}$ ($\Pi_{in}$) takes into account the cases that the SLE trace separates (does not separate) the twist operator from the interval $[x_1,x_2]\in\mathbb R$.
}
\end{figure}

In presence of the twist operator, $\Phi_{2,1}$, the partition function~\eqref{H0-weight} can be further decomposed into the weights representing possible configurations of the self-avoiding anchored path interacting with a twist defect. The possible configurations are shown in Fig.~\ref{P_gamma}. The corresponding statistical weights will be denoted as $\Pi_{in}$ and $\Pi_{out}$ representing the cases either the SLE trace separates or does not separate the twist operator from the interval $[x_1,x_2]\in\mathbb R$ correspondingly.
The statistical weights, $\Pi_{in}$ and $\Pi_{out}$, satisfy the following system of linear equations:
\begin{equation}\label{Pin-Pout}
\begin{aligned}
\Pi_{in}-\Pi_{out}&=H_1(z,\bar z;x_1,x_2)=H_0(x_1,x_2)\frac{2-\eta}{2\sqrt{1-\eta}},\\
\Pi_{in}+\Pi_{out}&=H_0(x_1,x_2)=(x_1-x_2)^{-5/4},
\end{aligned}
\end{equation}
where
\begin{equation}\label{eta-1point}
\eta=\frac{(z-\bar z)(x_1-x_2)}{(z-x_1)(\bar z-x_2)},
\end{equation}
and the coefficients of the linear combinations in front of the weights are determined as follows\footnote{These coefficients depend on $\kappa$. However, in the case $\kappa=8/3$ the coefficients are $\pm1$.}. To find the coefficient in front of $\Pi_{in}$ we send $z,\bar z\to x\in[x_1,x_2]$ on both sides of the first equation in~\eqref{Pin-Pout}. In this case $\eta=\epsilon(x_1-x_2)/(x-x_1)(x-x_2)$. Then, $\Pi_{in}\to H_{0}(x_1,x_2)$, $\Pi_{out}\to0$, and the expression on the right-hand side of~\eqref{Pin-Pout} goes to $H_0(x_1,x_2)$. Therefore, the coefficient in front of $\Pi_{in}$ equals $1$. Next, to find coefficient of $\Pi_{out}$ we send $z,\bar z\to x\in\mathbb R\setminus[x_1,x_2]$. In this limit we have $\Pi_{in}\to0$, $\Pi_{out}\to1$, while the expression on the right-hand side goes to $-1$, thus justifying the coefficient $-1$ in front of $\Pi_{out}$.

The system of equations~\eqref{Pin-Pout} determines probabilities for the SLE${}_{8/3}$ trace to wind in various ways about the point $z\in\mathbb H$. In particular, the probability that the SLE trace separates the point $z$ from the interval $[0,\infty]$, i.e., the left-crossing probability, $P_L(z)$, reads
\begin{equation}\label{Pleft-z}
P_{L}(z)=\frac{\Pi_{out}}{\Pi_{out}+\Pi_{in}}=\frac12-\frac{H_1(z,\bar z;x_1,x_2)}{2H_0(x_1,x_2)}=\frac12+\frac{\cos( \arg(z))}{2}.
\end{equation}
In the last equality we set $x_1=0$ and $x_2=\infty$, and took into account~\eqref{F1-int}. As expected, we obtain Schramm's formula~\eqref{P1-8/3} for $\kappa=8/3$.

%%%%%%%%%%%%%%%%%%%%%%%%%%%%%%%%%%

\subsection{The 2-point SLE${}_{8/3}$ passage probabilities}\label{s:2point}

%%%%%%%%%%%%%%%%%%%%%%%%%%%%%%%%%%

We already noted that the SLE${}_{8/3}$ left-crossing probability is determined by the correlation function containing the twist operator, $\Phi_{2,1}$, and a pair of the boundary 1-leg operators, $\Phi_{1,2}$. Similarly, in presence of two marked points in the bulk, $z_1,z_2\in\mathbb H$, the passage probabilities are determined by the correlation functions of two twist operators at the points $z_1,z_2$, and a pair of the 1-leg boundary operators at the points $x_1,x_2\in\mathbb R$:
\begin{equation}\label{Pi2-def}
H_2(z_1,\bar z_1,z_2,\bar z_2;x_1,x_2)=\langle \Phi_{2,1}(z_1,\bar z_1)\Phi_{2,1}(z_2,\bar z_2)\Phi_{1,2}(x_1)\Phi_{1,2}(x_2) \rangle_{\mathbb H}.
\end{equation}
The correlation function~\eqref{Pi2-def} is specified by the boundary conditions, which determine an appropriate linear combination of two \textit{conformal blocks} that contribute to the correlation function.

Before defining conformal blocks, it is convenient to recast the correlation function in $\mathbb H$ into the correlation function in $\mathbb C$. By replacing the antiholomorphic coordinates, $\bar z_1,\bar z_2\in\mathbb H$, by the holomorphic coordinates, $z_1^*,z_2^*\in\mathbb C$, we consider the 6-point correlation function
\begin{equation}\label{H2*}
H_2(z_1,z_1^*,z_2,z_2^*;x_1,x_2)
=\langle\prod_{i=1}^2\Phi_{2,1}(z_i) \Phi_{2,1}(z_i^*)\Phi_{1,2}(x_1) \Phi_{1,2}(x_2) \rangle.
\end{equation}
The conformal symmetry implies that it can be written in the form:
\begin{equation}
H_2(z_1,z_1^*,z_2,z_2^*;x_1,x_2)=\frac{G_2(\eta_1,\eta_2,\eta_3)}{(x_2-x_1)^{2h_{1,2}}(z_1- z_1^*)^{2h_{2,1}}(z_2-z_2^*)^{2h_{2,1}}}.
\end{equation}
Here $G_2(\eta_1,\eta_2,\eta_3)$ is a function of the cross-ratios $\eta_1=\eta(z^*_1)$, $\eta_2=\eta(z_2)$, and $\eta_3=\eta(z_2^*)$, where
\begin{equation}\label{eta-def}
\eta(s) = \frac{(z_1 - s)(x_1 - x_2)}{(z_1 - x_1)(s-x_2)}.
\end{equation}
In particular, $\eta(s)=1-s/z_1$ when $x_1=0$, and $x_2\to\infty$. One can also consider the correlation function $H_{2}(0,\eta_1,\eta_2,\eta_3;1,\infty)$ which can be written in the form
\begin{equation}\label{Pi2-eta}
H_{2}(0,\eta_1,\eta_2,\eta_3;1,\infty)=\frac{G_2(\eta_1,\eta_2,\eta_3)}{\eta_1^{2h_{2,1}}(\eta_2-\eta_1)^{2h_{2,1}}}.
\end{equation}
By eliminating the function $G_2$ from eqs.~\eqref{Pi2-def} and~\eqref{Pi2-eta} we obtain the following relation for the correlation function $H_2$:
\begin{equation}\label{F2kappa-Phi}
H_{2}(z_1,z_1^*,z_2,z_2^*;x_1,x_2)=\frac{\eta_1^{2h_{2,1}}(\eta_2-\eta_1)^{2h_{2,1}} H_{2}(0,\eta_1,\eta_2,\eta_3;1,\infty)}{(x_1-x_2)^{2h_{1,2}}(z_1- z_1^*)^{2h_{2,1}}(z_2-z_2^*)^{2h_{2,1}}}.
\end{equation}

The null state conditions for the operators $\Phi_{1,2}$ and $\Phi_{2,1}$ lead to six PDEs for $G_2(\eta_1,\eta_2,\eta_3)$. The required solution to these equations must satisfy the following limiting condition:
\begin{equation}\label{H2lim}
\lim_{z_1-z_2\to \infty}\frac{H_2(z_1,\bar z_1,z_2,\bar z_2,x_2,x_2)}{H_0(x_1,x_2)} = H_1(z_1,\bar z_1,x_1,x_2)H_1(z_2,\bar z_2,x_1,x_2).
\end{equation}
In Ref.~\cite{SC09} Simmons and Cardy proposed a unique solution to the system of the equations, which satisfies the limiting condition~\eqref{H2lim}. However, in the case of $N\geq 3$ twist operators in the bulk the system of the null-state PDEs is very difficult to solve directly.

Below, we obtain Dotsenko-Fateev representation for the correlation function containing two twist operators. Following the guideline described in the previous section we consider the following product of the vertex operators $V_{2,1}$, $V_{1,2}$ and $V_{-1,-2}$:
\begin{multline}\label{F2}
\mathcal F_2(0,\eta_1,\eta_2,\eta_3;1,\infty;\gamma_1,\gamma_2)= \\
= \langle V_{2,1}(0)\prod_{i=1}^3V_{2,1}(\eta_i)V_{1,2}(1)V_{-1,-2}(\infty)(Q^-)^2\rangle,
\end{multline}
where we inserted two screening charges $Q^-$ inside the correlation function of vertex operators in order to satisfy the neutrality condition~\eqref{neutrality}. 

We will refer to the correlation function~\eqref{F2} as to the \textit{conformal block}. It depends on the integration contours, $\gamma_1$ and $\gamma_2$, which determine the screening charges~\eqref{Qpm}. The correlation function of primary fields~\eqref{H2*} is given by an appropriate linear combination of these blocks:
\begin{equation}\label{H2-X}
H_2(0,\eta_1,\eta_2,\eta_3;1,\infty)=\sum_{i,j}N(\gamma_i,\gamma_j)\mathcal F_2(0,\eta_1,\eta_2,\eta_3;1,\infty;\gamma_i,\gamma_j),
\end{equation}
where the coefficients $N(\gamma_i,\gamma_j)$ depends on the boundary conditions. In the case of the 6-points function~\eqref{F2} there exist 10 natural couples of the contours ($\gamma_i,\gamma_j$). However, one can argue that only one choice of the contours is relevant, namely, $\gamma_1$ and $\gamma_2$ are simple paths connecting $0,\eta_1$ and $\eta_2,\eta_3$ respectively. Below, we present simple reasoning in support of the statement. Note, however, that our conclusion is justified by explicit calculations~\cite{SC09}.

\begin{figure}[t]
\centering
\includegraphics[width=.55\columnwidth]{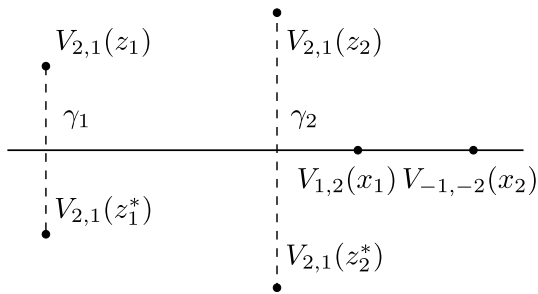}
\caption{\label{cf_real} The conformal block $\mathcal F_2(z_1,z_2^*,z_2,z_2^*;x_1,x_2)$ is shown with respect to the bulk-boundary fusion. The boundary is shown by the solid line. The dashed lines represent the integration paths for two screening charges $Q^-$.
}
\end{figure}

One can show, that the \textit{bulk-boundary} fusion, $\Phi_{2,1}(z_2)\Phi_{2,1}(z_2^*)$ as $z_2,z_2^*\to x\in\mathbb R$, can be realized via the identity channel only~\cite{SC09} (we briefly discuss the algebraic structure of $c=0$ LCFT in Section~\ref{s:Green}). There exist one conformal block which satisfy this requirement, namely, $\mathcal F_2(z_1,z_1^*,z_2,z_2^*;x_1,x_2)$ with the integration path connecting $z_2,z_2^*$. This can be easily shown by inserting the product $\int_{z_2}^{z_2^*}V_{2,1}(z_2)V_{2,1}(z_2^*)V_{-}(u)\,du$ into the conformal block~\eqref{F1-def}. The choice of the integration contour (from $z_2$ to $z_2^*$) implies that the pair of twist operators, $V_{2,1}(z_2)$ and $V_{2,1}(z_2^*)$, fuse only via the identity channel when $z_2,z_2^*\to x\in\mathbb R$. Indeed, by fusing these operators we obtain the vertex operator $V_{\alpha}(x)$ with the charge $\alpha=2\alpha_{2,1}=\alpha_{3,1}$. By adding the screening charge $\alpha_-$, the total charge vanishes, $2\alpha_{2,1}+\alpha_-=0$, implying that the result is the identity operator. However, the screening charge is pulled in with the vertex operators only if the path $\gamma_2$ contracts to a point in this process. Hence, the contour $\gamma_2$ connects $z_2$ and $z_2^*$. Similarly, one can prove that $\gamma_1$ connects $z_1$ and $z_1^*$. We show the conformal block $\mathcal F_2(z_1,z_1^*,z_2,z_2^*;x_1,x_2)$ in Fig.~\ref{cf_real}. Under M\"obius transformation~\eqref{eta-def} these contours become $\gamma(z_1,z_1^*)\to \gamma(0,\eta_1)$ and $\gamma(z_2,z_2^*)\to \gamma(\eta_2,\eta_3)$ as proposed below eq.~\eqref{H2-X}.

By using~\eqref{VV-correlation} and evaluating the conformal block~\eqref{F2} we obtain the integral representation of the correlation function\footnote{It is easy to see, that the conformal block shown in Fig.~\ref{cf_real} is a real function. Hence, the correlation function~\eqref{F2kappa-Phi} is also real.}:
\begin{equation}\label{H2-int-repr}
H_2(0,\eta_1,\eta_2,\eta_3;1,\infty)=N_2\prod_{i=1}^3\frac{\eta_i^{\kappa/8}}{(1-\eta_i)^{1/2}}\prod_{i<j}(\eta_i-\eta_j)^{\kappa/8}\mathcal J_2(\eta_1,\eta_2,\eta_3),
\end{equation}
where $N_2$ is the normalization constant, and $\mathcal J_2(\eta_1,\eta_2,\eta_3)$ denotes the following double contour integral:
\begin{multline}\label{mathcalJ-def}
\mathcal J_2(\eta_1,\eta_2,\eta_3)=\int_{\eta_2}^{\eta_3} du_1\int_{0}^{\eta_1}du_2 (u_1-u_2)^{\kappa/2}\times\\
\times\prod_{i=1}^2 u_i^{-\kappa/4}(1-u_i)\prod_{j=1}^3(u_i-\eta_j)^{-\kappa/4}.
\end{multline}
The normalization constant is determined by sending $\eta_3\to \eta_1$, so that the correlation function $H_2$ reduces to $H_1$. Simple calculation shows $N_2=N_1^2$.

%Let us consider the integral $\mathcal J_2(\eta_1,\eta_2,\eta_3)$ in the limit $\eta_2\to\eta_3$. We set $\eta_2-\eta_3=\epsilon$, and consider the leading term of $\mathcal J_2$ of the small-$\epsilon$ expansion:
%\begin{multline}\label{J2-expansion}
% \lim_{\epsilon\to0}\mathcal J_2(\eta_1,\eta_2,\eta_2+\epsilon)=\epsilon^{1-\kappa/2}\mathcal J_1(\eta_1)\eta_2^{-\kappa/4}(1-\eta_2)(\eta_1-\eta_2)^{-\kappa/4}\times\\
%\times\oint_{\gamma(0,1)}du u^{-\kappa/4}(1-u)^{-\kappa/4}+O(\epsilon^{2-\kappa/2}),
%\end{multline}
%where $\mathcal J_1(\eta)$ is given by~\eqref{J1-def}. From~\eqref{J2-expansion} the small-$\epsilon$ expansion of the correlation function follows,
%\begin{equation}
% \lim_{\epsilon\to0}H_2(0,\eta_1,\eta_2,\eta_2+\epsilon;1,\infty)=H_1(\eta_1;1,\infty)+O(\epsilon),
%\end{equation}
%where the function $H_1$ is given by the right-hand side of~\eqref{H1*-int}, and the normalization constant $N_2=N_1^2$ (see eq.~\eqref{N1-def}).

\begin{figure}[t]
\centering
\includegraphics[width=1\columnwidth]{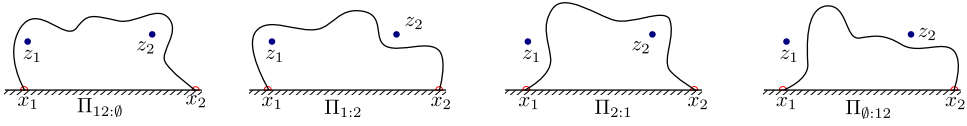}
\caption{\label{P4_gamma} The four weights of the upper-half plane SLE${}_{8/3}$ with respect to the pair of twist defects at the points $z_1$ and $z_2$. The weight $\Pi_{ij:kl}$ counts the cases when SLE trace does not separates the twist operators at the points $z_i$, $z_j$ from the interval $[x_1,x_2]\in\mathbb R$, while the points $z_k$, $z_l$ are separated from the interval.
}
\end{figure}

Now we are ready to determine the probabilities for the SLE${}_{8/3}$ trace to wind about 2 marked points in $\mathbb H$. In presence of twist defects the partition function~\eqref{H0-weight} for SLE from $x_1$ to $x_2$ can be decomposed into the sum of the weights depending on the winding of the paths around two points. We label these weights by $\Pi_{12:\emptyset}$, $\Pi_{1:2}$, $\Pi_{2:1}$, and $\Pi_{\emptyset:12}$, where $\Pi_{ij:kl}$ denotes the weight of the paths that separate the points $z_k$, $z_l$ from the interval $[x_1,x_2]\in\mathbb R$, while the points $z_i$, $z_j$ remain unseparated (see Fig.~\ref{P4_gamma}). Following the guideline described in the previous section we decompose the correlation functions $H_0$, $H_1$, and $H_2$ in terms of the weights of trace configurations:
\begin{equation}\label{Pi2-dec}
\begin{aligned}
\Pi_{12:\emptyset}+\Pi_{1:2}+\Pi_{2:1}+\Pi_{\emptyset:12} & =H_0(x_1,x_2),\\
\Pi_{12:\emptyset}+\Pi_{1:2}-\Pi_{2:1}-\Pi_{\emptyset:12} & = H_1(z_1,\bar z_1;x_1,x_2),\\
\Pi_{12:\emptyset}-\Pi_{1:2}+\Pi_{2:1}-\Pi_{\emptyset:12} & = H_1(z_2,\bar z_2;x_1,x_2),\\
\Pi_{12:\emptyset}-\Pi_{1:2}-\Pi_{2:1}+\Pi_{\emptyset:12} & = H_2(z_1,\bar z_1,z_2,\bar z_2;x_1,x_2).
\end{aligned}
\end{equation}
The coefficients of these linear combinations in front of the weights are determined similarly to the previous case (see the discussion below Eq.~\eqref{Pin-Pout}, and Ref.~\cite{SC09}). By solving these equations, one obtains the weights of various trace configurations, e.g.,
\begin{multline}\label{prob-2}
% \Pi_{12:\emptyset}&=\frac{1}{4}(H_0(x_1,x_2)+H_1(z_1,\bar z_1,x_1,x_2) + H_1(z_2,\bar z_2,x_1,x_2) + H_2(z_1,\bar z_1,z_2,\bar z_2,x_1,x_2)),\\
% \Pi_{1:2}&=\frac{1}{4}(H_0(x_1,x_2)+H_1(z_1,\bar z_1,x_1,x_2) - H_1(z_2,\bar z_2,x_1,x_2) - H_2(z_1,\bar z_1,z_2,\bar z_2,x_1,x_2)),\\
% \Pi_{2:1}&=\frac{1}{4}(H_0(x_1,x_2)-H_1(z_1,\bar z_1,x_1,x_2) + H_1(z_2,\bar z_2,x_1,x_2) - H_2(z_1,\bar z_1,z_2,\bar z_2,x_1,x_2)),\\
\Pi_{\emptyset:12}=\frac{1}{4}(H_0(x_1,x_2) - H_1(z_1,\bar z_1;x_1,x_2) - \\
- H_1(z_2,\bar z_2;x_1,x_2) + H_2(z_1,\bar z_1,z_2,\bar z_2;x_1,x_2)),
\end{multline}
and the probabilities of the corresponding events read
\begin{equation}\label{P2-prob}
P_{ij:kl}=\frac{\Pi_{ij:kl}}{\Pi_{12:\emptyset}+\Pi_{1:2}+\Pi_{2:1}+\Pi_{\emptyset:12}}=\frac{\Pi_{ij:kl}}{H_0(x_1,x_2)}.
\end{equation}
In particular, the probability, $P_{L}(z_1,z_2)$, that the SLE trace passes to the left of both points $z_1,z_2\in\mathbb H$ is determined by~\eqref{prob-2}. Taking into account~\eqref{Pleft-z} we obtain
\begin{equation}
P_L(z_1,z_2)=\frac12 P_L(z_1)+\frac12 P_L(z_2)+\frac{1}{4}\left(\frac{H_2}{H_0}-1\right).
\end{equation}

%%%%%%%%%%%%%%%%%%%%%%%%%%%%%%%%%%

\subsection{The $N$-point SLE${}_{8/3}$ passage probabilities}\label{s:Npoint}

%%%%%%%%%%%%%%%%%%%%%%%%%%%%%%%%%%

In this section we briefly summarize the result for probabilities of the SLE${}_{8/3}$ trace to wind about $N$ points in $\mathbb H$. First, we define the following set of the multi-point correlation functions:
\begin{equation}\label{Hn-def}
H_n(z_1,\bar z_1,\dotsc,z_n,\bar z_n;x_1,x_2)=\langle \prod_{i=1}^n\Phi_{2,1}(z_i,\bar z_i) \Phi_{1,2}(x_1)\Phi_{1,2}(x_2) \rangle_{\mathbb H},
\end{equation}
for $n=0,1,\dotsc,N$. In order to obtain the Coulomb gas representation for the correlation functions~\eqref{Hn-def} we consider the product of $2n$ vertex operators $V_{2,1}(z_i)$, $V_{2,1}(z_i^*)$, $i=1,2,\dotsc,n$, and the boundary operators, $V_{1,2}(x_1)$ and $V_{-1,-2}(x_2)$ at $x_1,x_2\in\mathbb R$. The correlation function of vertex operators,
\begin{equation}\label{Fn-cgas}
\mathcal F_n(z)=\langle \prod_{i=1}^{n} V_{2,1}(z_i)V_{2,1}(z^*_i)V_{1,2}(x_1)V_{-1,-2}(x_2)(Q^-)^n\rangle,
\end{equation}
requires $n$ screening charges in order to satisfy the neutrality condition~\eqref{neutrality}. We choose the integration contours in~\eqref{Fn-cgas} to be simple curves connecting the points, $z_i,z_i^*$, $i=1,2,\dotsc, n$, pairwise. In this case the bulk-boundary fusion of the vertex operators, $V_{2,1}(z_i)$ and $V_{2,1}(z_i^*)$ as $z_i^*,z_i\to x\in\mathbb R$, is realized via the identity channel (see the discussion below Eq.~\eqref{H2-X}). Hence, we obtain the following Coulomb gas representation for the function $H_n$:
\begin{multline}\label{Hn-c}
H_n(z_1,z_1^*,\dotsc,z_n,z_n^*;x_1,x_2)=\frac{N_n \prod_{i=1}^{2n-1}(\eta_i-\eta_{i-1})^{2h_{2,1}} }{(x_1-x_2)^{2h_{1,2}}\prod_{i=1}^n(z_i- z_i^*)^{2h_{2,1}}}\times\\
\times\prod_{i=1}^{2n-1}\eta_i^{\kappa/8}(1-\eta_i)^{-1/2}\prod_{k<l}(\eta_k-\eta_l)^{\kappa/8}\mathcal J_{n}(\eta_1,\dotsc,\eta_{2n-1}),
\end{multline}
where $N_n=N_1^n$ is the normalization constant, the function $\mathcal J_n$ is determined by the $n$-fold integral
\begin{multline}
\mathcal J_n(\eta_1,\dotsc,\eta_{2n-1}) = \int_{0}^{\eta_1}du_1 \int_{\eta_2}^{\eta_3}du_2 \cdots \int_{\eta_{2n-2}}^{\eta_{2n-1}}du_n\times\\
\times\prod_{i<j}(u_i-u_j)^{\kappa/2} \prod_{i=1}^n u_i^{-\kappa/4}(1-u_i)\prod_{k=1}^{2n-1}(u_i-\eta_k)^{-\kappa/4},
\end{multline}
and we introduced the cross-ratios, $\eta_i=\eta(u_i)$, $i=0,1,\dotsc,2n-1$, where $\eta(u)$ is specified by eq.~\eqref{eta-def}, and $u_i\in\{z_1,z_1^*,z_2,z_2^*,\dotsc,z_n,z_n^*\}$. Below, we suppose that $\kappa=8/3$.

The correlation functions~\eqref{Hn-def} determine the probabilities of the SLE${}_{8/3}$ trace to wind about the points $z_1,z_2,\dotsc,z_n\in\mathbb H$ in what follows. In presence of $n$ twist operators the partition function of the SLE${}_{8/3}$ curve can be decomposed into the sum of the weights depending on the winding of the curve about the points. In order to label these weights, we introduce the following notation. Let $I_n=\{1,2,\dotsc,n\}$ be the set of $n$ integers, and $I_n\subset I_N$. We decompose the set $I_n$ into two subsets, $I_n^+$ and $I_n^-$, so that $I_n^+\cup I_n^-=I_n$, $I_n^+\cap I_n^- = \emptyset$. Besides, let us introduce the set of points, $Z_{I}=\{z_i|i\in I\}$, labeled by the integers from the set $I$. By $\Pi_{I_n^+:I_n^-}$ we denote the weight of the SLE${}_{8/3}$ traces which separates the points $Z_{I^-_n}$ away from the interval $[x_1,x_2]\subset\mathbb R$, while the points $Z_{I^+_n}$ remain unseparated. Then, we can decompose the partition function in terms of the weights as follows (c.f., eqs.~\eqref{Pi2-dec}):
\begin{equation}\label{Pi-N}
\sum_{I_N^++I_N^-=I_N}(-1)^{\# I_n^-}\Pi_{I_N^+:I_N^-}=H_n(Z_{I_n};x_1,x_2),
\end{equation}
for $n=0,1,\dotsc,N$. In~\eqref{Pi-N} the sum is taken over all decompositions of $I_N$ into two subsets $I_N^+$ and $I_N^-$. The coefficients $(-1)^{\# I_n^-}$ can be obtained by sending $z_i,\bar z_i\to x\in[x_1,x_2]\ (x\in\mathbb R\setminus[x_1,x_2])$ as explained in the discussion below Eq.~\eqref{Pin-Pout}. Therefore, we obtain $2^N$ linear equations for the $2^N$ unknown statistical weights $\Pi_{I_N^+:I_N^-}$. 

Eqs.~\eqref{Pi-N} allow us to determine statistical weights of SLE${}_{8/3}$ traces from $x_1$ to $x_2$ to pass to the right of the points $Z_{I^-_N}$ and to the left of the points $Z_{I^+_N}$ (c.f., Eqs.~\eqref{prob-2}):
\begin{equation}
\Pi_{I_N^+:I_N^-} = 2^{-N}\sum_{n=0}^N \sum_{I_n} (-1)^{\#(I_N^-\cap I_n)}H_n(Z_{I_n};x_1,x_2),
\end{equation}
and the probability that the SLE${}_{8/3}$ trace from $x_1$ to $x_2$ separates the points $Z_{I_N^-}$ from the interval $[x_1,x_2]$ reads
\begin{equation}
P_{I_N^+:I_N^-} = \frac{\Pi_{I_N^+:I_N^-}}{H_0(x_1,x_2)}.
\end{equation}

%%%%%%%%%%%%%%%%%%%%%%%%%%%%%%%%%

\section{Green's functions for SLE${}_{8/3}$}\label{s:Green}

%%%%%%%%%%%%%%%%%%%%%%%%%%%%%%%%%

%%%%%%%%%%%%%%%%%%%%%%%%%%%%%%%%%

\subsection{The 1-point SLE${}_{8/3}$ Green's function}

%%%%%%%%%%%%%%%%%%%%%%%%%%%%%%%%%

In this section we discuss the probability that the SLE${}_{8/3}$ trace passes in the $\epsilon$-neighborhood of the marked point. This probability is closely related with the one-point SLE${}_{8/3}$ Green's function. More specifically, consider the SLE${}_{8/3}$ trace from $x_1$ to $x_2$. Then, the probability, $\mathbb P\{z<\epsilon;x_1,x_2\}$, that the trace passes in the $\epsilon$-neighborhood of the point $z\in\mathbb H$ vanishes as follows:
\begin{equation}\label{G-def}
\lim_{\epsilon\to0} \epsilon^{-2/3}\mathbb P\{z<\epsilon;x_1,x_2\}=c G^\text{SLE}_{\mathbb H}(z;x_1,x_2)+O(\epsilon),
\end{equation}
where $c$ is a constant, and $G^\text{SLE}_{\mathbb H}(z;x_1,x_2)$ is called the one-point Green's function of the SLE trace.

In order to evaluate the one-point Green's function we use the results obtained in the previous section. We consider the probability that the SLE${}_{8/3}$ trace passes between the points $z_1,z_2\in\mathbb H$. There are two possible trace configurations, which contribute to the probability, namely, $\Pi_{1:2}$ and $\Pi_{2:1}$ (see Fig.~\ref{P4_gamma}). Hence, the probability of the event is given by
\begin{multline}\label{P-12}
P(z_1,z_2;x_1,x_2)=\frac{\Pi_{1:2}+\Pi_{2:1}}{\Pi_{12:\emptyset}+\Pi_{1:2}+\Pi_{2:1}+\Pi_{\emptyset:12}}= \\ 
=\frac12-\frac{H_2(z_1,\bar z_1,z_2,\bar z_2;x_1,x_2)}{2H_0(x_1,x_2)}.
\end{multline}
Further, we set $z_1=z+\epsilon \nu/2$, $z_2=z-\epsilon \nu/2$, where $z\in\mathbb H$, $\epsilon\ll1$, $|\nu|=1$, and consider the series expansion of $P(z_1,z_2;x_1,x_2)$ in the limit $\epsilon\to0$. The leading term of the small-$\epsilon$ expansion determines the Green's function of the trace as follows:
\begin{equation}\label{G1-lim-def}
\lim_{\epsilon\to0} \epsilon^{-2/3}P\left(z-\frac{\epsilon\nu}{2},z+\frac{\epsilon\nu}{2};x_1,x_2\right)= c_1 G_{\mathbb H}(z;x_1,x_2),
\end{equation}
where $c_1$ is a constant.

In order to evaluate~\eqref{G1-lim-def} one needs to study the series expansion of the 4-point correlation function~\eqref{Pi2-def} as $z_1\to z_2$. In the CFT framework the required expansion can be obtained by using the so-called \textit{operator product expansion} (OPE) of the primary fields $\Phi_{2,1}(z_1)\Phi_{2,1}(z_2)$ inside the correlation function. The form of the OPE can be deduced from the global conformal invariance including the form of the two- and three-point functions and their symmetry properties\footnote{This form of the OPE is typical for rational CFTs, while in LCFTs the OPE of certain operators can be modified.}:
\begin{equation}\label{OPE}
\Phi_{h_i}(z)\Phi_{h_j}(0)=z^{h_k-h_i-h_j}\sum_k C^{k}_{i,j}\left(\Phi_{h_k}(0)+\sum_{\{n\}}\beta_{i,j}^{k,\{n\}} z^{|\{n\}|}\Phi_{h_k}^{(-\{n\})}(0)\right),
\end{equation}
where the coefficients $\beta_{i,j}^{k,\{n\}}$ are fixed by conformal invariance, and $\Phi_{h_k}^{(-\{n\})}$ denotes the contribution of the $|\{n\}|$-level descendant operators:
\begin{equation}
\Phi_{h_k}^{(-\{n\})}=L_{-n_1}L_{-n_2}\cdots L_{-n_l}\Phi_{h_k},
\end{equation}
where $L_{-n}$ are the generators of the Virasoro algebra~\cite{BPZ}. The structure constants $C_{ij}^k$ are determined by the two- and three-point functions\footnote{In LCFT certain structure constants become functions containing logarithms.},
\begin{equation}
C_{ij}^k=\lim_{z\to\infty}|z|^{4h_i}\langle\Phi_{h_i}(z,\bar z)\Phi_{h_j}(1)\Phi_{k_k}(0)\rangle,
\end{equation}
where the normalization $\langle\Phi_{h_i}(z,\bar z)\Phi_{h_i}(0)\rangle=|z|^{-2h_i}$ is assumed. Note, that the structure constants are not fixed by conformal invariance, and the additional constraints follow from the request of associativity of the operator algebra~\cite{BPZ}. However, once the structure constants are known all correlation functions can be in principle computed.

Let us briefly recall the structure of $c=0$ LCFT (see Refs.~\cite{MR07,MR08} for details\footnote{Note, that we follow the notation of Ref.~\cite{SC09}, so that the K\^ac indices are in reverse order to those in~\cite{MR07,MR08}}.). By $\mathcal V_{r,s}$ we denote the Verma module generated from the state $|\Phi_{r,s}\rangle$. In $c=0$ LCFT the vacuum module is indecomposable $\mathcal M_{1,1}=\mathcal V_{1,1}/\mathcal V_{4,1}$. Furthermore, the physical module corresponding to the primary field $\Phi_{2,1}$ is $\mathcal M_{2,1}=\mathcal V_{2,1}/\mathcal V_{5,1}$. The fusion of this modules with itself reads:
\begin{equation}\label{M21M21}
\mathcal M_{2,1}\times \mathcal M_{2,1} = \mathcal M_{1,1} + \mathcal M_{3,1},
\end{equation}
where $\mathcal M_{1,1}$ was introduced earlier, and $\mathcal M_{3,1}$ turns out to be the irreducible module with $h_{3,1}=1/3$. The fusion rule~\eqref{M21M21} implies the following form of the OPE of primary fields $\Phi_{1,1}(\epsilon)$ and $\Phi_{3,1}(0)$:
\begin{equation}\label{OPE-21}
\lim_{\epsilon\to0}\Phi_{2,1}(\epsilon)\Phi_{2,1}(0)=\Phi_{1,1}(0)+C_{3,1}\epsilon^{1/3} \Phi_{3,1}(0)+O(\epsilon),
\end{equation}
where $C_{3,1}$ is a fixed OPE coefficient. In this case one can say that the OPE is realized via two \textit{channels}: the first one involves $\Phi_{1,1}$, while the second one involves $\Phi_{3,1}$. However, in the boundary CFT the general form of the OPE can be modified because of the boundary conditions. As we will see below, the case of $c=0$ boundary LCFT is even more tricky.

Let us consider the correlation function~\eqref{H2*}, and examine the \textit{bulk-boundary} fusion $\Phi_{2,1}\Phi_{2,1}$. The coulomb gas representation~\eqref{H2-int-repr} allows us to obtain the OPE explicitly. By shrinking the integration contour connecting $z_1,\bar z_1$, one obtains the small-$\epsilon$ expansion of $H_2(x+\epsilon/2,x-\epsilon/2,z_2,\bar z_2,x_1,x_2)$, where $x\in\mathbb R$. It has the form $g_0+\epsilon g_1+\epsilon^2 g_2+\dotsc$, where $g_n$ with $n\geq0$ are certain functions of the coordinates $\{x,z_2,\bar z_2,x_1,x_2\}$. By comparing this expansion with OPE~\eqref{OPE-21} we conclude that the bulk-boundary fusion is realized via the identity fusion channel, while the second channel, $\Phi_{3,1}$, is forbidden. This result can be also justified by computations in Ref.~\cite{SC09}.

\begin{figure}[t]
\centering
\includegraphics[width=.7\columnwidth]{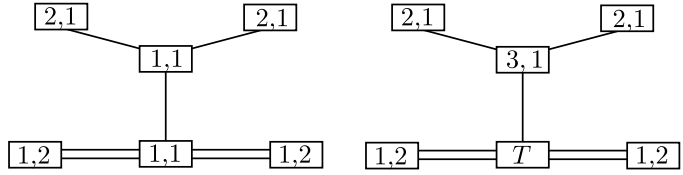}
\caption{\label{c-blocks}
Two conformal blocks, which contribute to the correlation function $H_2(z_1,\bar z_1,z_2,\bar z_2;x_1,x_2)$, are shown with respect to the \textit{bulk-bulk} fusion, $\Phi_{2,1}\Phi_{2,1}$. The fusion can be realized via two channels: $\Phi_{1,1}$ and $\Phi_{3,1}$. By $T$ we denote the stress-energy tensor, and rectangles $[m,n]$ correspond to the fields $\Phi_{m,n}$. Note, that the boundary operators are connected by double lines.
}
\end{figure}

Further, we consider the \textit{bulk-bulk} fusion of the fields $\Phi_{2,1}\Phi_{2,1}$ inside the correlation function as $z_2\to z_1$, $\bar z_2\to \bar z_1$. Explicit calculations shows that both channels, $\Phi_{1,1}$ and $\Phi_{3,1}$, appear in this case~\cite{SC09}. Hence, the small-$\epsilon$ expansion of the correlation function $H_2$ containing two twist operators can be obtained from~\eqref{OPE-21}. It reads
\begin{multline}\label{H2-expansion}
\lim_{\epsilon\to0}H_2(z+\epsilon \nu/2, z-\epsilon w/2,\bar z+\epsilon \bar \nu/2,\bar z-\epsilon \bar w/2;x_1,x_2)=H_0(x_1,x_2)+\\
+(C_{3,1})^2\epsilon^{2/3}\langle \Phi_{3,1}(z,\bar z)\Phi_{1,2}(x_1)\Phi_{1,2}(x_2)\rangle_{\mathbb H}+O(\epsilon).
\end{multline}
By using~\eqref{H2-expansion} we recast the probability~\eqref{P-12} in the form
\begin{equation}\label{P-12-G}
\lim_{\epsilon\to0} \epsilon^{-2/3}P(z-\epsilon\eta,z+\epsilon\eta;x_1,x_2)=(C_{3,1})^2 G_{\mathbb H}(z;x_1,x_2),
\end{equation}
where the Green's function (compare~\eqref{P-12-G} with~\eqref{G-def}) is written in terms of the following correlation function:
\begin{equation}\label{Pz-epsilon}
G_{\mathbb H}(z;x_1,x_2)=-\frac{ \langle\Phi_{3,1}(z,\bar z)\Phi_{1,2}(x_1)\Phi_{1,2}(x_2)\rangle_{\mathbb H}}{2H_0(x_1,x_2)}.
\end{equation}

Hence, we proposed an explicit expression of the one-point SLE${}_{8/3}$ Green's function in terms of the correlation function of the field $\Phi_{3,1}$ in the bulk, and two fields $\Phi_{1,2}$ at the boundary in $c=0$ LCFT. Recall, that the SLE${}_{8/3}$ Green's function defined by~\eqref{G-def} can be evaluated explicitly~\cite{RS05},
\begin{equation}\label{G-SLE}
G_{\mathbb H}^{\text{SLE}_{8/3}}(z;0,\infty)=(\Im z)^{-2/3}\sin^2(\arg(z)).
\end{equation}
In the next section we will show that $G_{\mathbb H}(z;0,\infty)=G_{\mathbb H}^{\text{SLE}_{8/3}}(z;0,\infty)$ by obtaining an explicit Coulomb gas representation for the correlation function on the right-hand side of~\eqref{G-SLE}. %For $\kappa=8/3$ the Coulomb gas integral reduces to the elementary functions.

\subsection{Coulomb gas representation of the Green's function}

In this section we obtain a Coulomb gas representation for the one-point SLE${}_{8/3}$ Green's function. The correlation function on the right-hand side of eq.~\eqref{Pz-epsilon} can be realized via the correlation functions of vertex operators containing a single screening charge $Q^+$:
\begin{equation}\label{2Phi31}
\langle\Phi_{3,1}(z)\Phi_{3,1}(z^*)\Phi_{1,2}(x_1)\Phi_{1,2}(x_2)\rangle=\sum_{\{\gamma\}} M(\gamma)\mathcal H_1(z,z^*;x_1,x_2;\gamma),
\end{equation}
where the conformal blocks, $\mathcal H_1$, are given by
\begin{equation}\label{H1-def}
\mathcal H_1(z, z^*;x_1,x_2;\gamma)=\langle V_{3,1}(z)V_{-3,-1} (z^*)V_{1,2}(x_1) V_{1,2}(x_2) Q^+\rangle,
\end{equation}
and the coefficients, $M(\gamma)$, depend on the integration contour. By taking into account~\eqref{VV-correlation} we obtain the following integral representation for the conformal block:
\begin{equation}
\mathcal H_1(z,z^*;x_1,x_2;\gamma)=\frac{\int_{\gamma} du\, u^{8/\kappa-4}(1-u)^{-4/\kappa}(u-\eta)^2}{(z-z^*)^{2h_{3,1}}(x_1-x_2)^{2h_{1,2}}(1-\eta)},
\end{equation}
where $\eta$ is the standard cross-ratio introduced earlier~\eqref{G1}.

One can argue, that in the case $\kappa=8/3$ only one conformal block (with respect to the bulk-boundary fusion) contribute to the correlation function~\eqref{2Phi31}. This block is specified by the integration path connecting $0$ and $\eta$. Indeed, explicit calculation shows that the bulk-boundary fusion $\Phi_{3,1}\Phi_{3,1}$ is realized via the weight 2 operator, namely, the stress-energy tensor~\cite{SC09}. When $z,z^*\to x\in\mathbb R$ the operators, $V_{3,1}(z)$ and $V_{-3,-1}(z^*)$, fuse to $V_\alpha(x)$ with $\alpha=2\alpha_0$. By adding the screening charge $\alpha_+$, we determine the total charge $2\alpha_0+\alpha_+$, and the conformal dimension of the operator is equal to $h_{2\alpha_0+\alpha_+}=2$, i.e., the dimension of the stress-energy tensor. However, the screening charge is pulled in with the fusion only if the contour $\gamma$ contracts to a point in this process. Hence, $\gamma$ is a simple path connecting $z$ and $z^*$.

Below, we suppose that $\kappa=8/3$. After substituting $u=\eta t$ in the integrand of $\mathcal H_1(z,z^*;x_1,x_2;\gamma)$, and taking the integration contour to be a simple path connecting 0 and $\eta$, we obtain the following representation for correlation function~\eqref{2Phi31}, namely,
\begin{equation}\label{Phi31-cf}
\langle\Phi_{3,1}(z,\bar z)\Phi_{1,2}(x_1)\Phi_{1,2}(x_2)\rangle_{\mathbb H}=\frac{M_1\eta^2}{(z-\bar z)^{2h_{3,1}}(x_1-x_2)^{2h_{1,2}}(1-\eta)},
\end{equation}
where $M_1$ is a normalization constant, and we set $z^*=\bar z$ in order to obtain the correlation function in $\mathbb H$. Thus, we find an explicit expression for the Green's function~\eqref{Pz-epsilon}:
\begin{equation}
G_{\mathbb H}(z;x_1,x_2) =\frac{1}{4(\Im z)^{2/3}}\frac{\eta^2}{1-\eta}.
\end{equation}
In the last expression we set $M_1=2^{-1/3}e^{2i\pi/3}$. By setting $x_1=0$, $x_2\to\infty$, so that $\eta=1-e^{-2i\arg(z)}$, we conclude, that the function~\eqref{Pz-epsilon} takes the form of the one-point Green's function for SLE${}_{8/3}$ in the upper-half plane~\eqref{G-SLE}, i.e., $G_{\mathbb H}(z;0,\infty)=G^{\text{SLE}_{8/3}}_{\mathbb H}(z;0,\infty)$. 

%%%%%%%%%%%%%%%%%%%%%%%%%%%%%%%%%%%

\subsection{The two-point Green's function}

%%%%%%%%%%%%%%%%%%%%%%%%%%%%%%%%%%%

In this section we outline a derivation of the two-point SLE${}_{8/3}$ Green's function. We use the results of the previous sections, where the probabilities for the SLE trace to wind in various ways about 4 marked points, $z_1,z_2,z_3$, and $z_4$, were obtained. In particular, the probability that the curve passes between the points $z_1,z_2$ and $z_3,z_4$ correspondingly is given by the normalized linear combination of 4 trace configurations\footnote{We denote the function $H_n(z_1,\bar z_1,\dotsc,z_n,\bar z_n;x_1,x_2)$ as $H_n(z_1,\dotsc,z_n;x_1,x_2)$ for brevity.}:
\begin{multline}\label{P12}
P(z_1,z_2,z_4,z_4;x_1,x_2)=\frac{\Pi_{13:24}+\Pi_{14:23}+\Pi_{23:14}+\Pi_{24:13}}{H_0(x_1,x_2)}=\\
=\frac14-\frac{H_{2}(z_1,z_2;x_1,x_2)}{4H_0(x_1,x_2)}- \frac{H_{2}(z_3,z_4;x_1,x_2)}{4H_0(x_1,x_2)}+\frac{H_{4}(z_1,z_2,z_3,z_4;x_1,x_2)}{4H_0(x_1,x_2)}.
\end{multline}
Here $H_n(z_1,\ldots,z_n;x_1,x_2)$ is the $n$-point correlation function in the upper-half plane~\eqref{Hn-def}. Let us set
\begin{equation}
\begin{gathered}
z_1=z+\epsilon \nu/2,\quad z_2=z-\epsilon \nu/2,\\
z_3=w+\delta \mu/2,\quad z_4=w - \delta\mu/2,
\end{gathered}
\end{equation}
where $z,w\in\mathbb H$, $\epsilon,\delta\ll1$, $|\nu|,|\mu|=1$, and consider the series expansion of the probability~\eqref{P12} in the limit $\epsilon,\delta\to0$. The leading term in the small-$\epsilon,\delta$ expansion determines the two-point SLE${}_{8/3}$ Green's function
\begin{multline}\label{zw-def}
\lim_{\epsilon,\delta\to0} \epsilon^{-2/3}\delta^{-2/3}P\left(z-\frac{\epsilon\nu}{2},z+\frac{\epsilon\nu}{2},w-\frac{\epsilon\mu}{2},w+\frac{\epsilon\mu}{2};x_1,x_2\right)=\\
=c_2 G_{\mathbb H}(z,w;x_1,x_2),
\end{multline}
where $c_2$ is a constant.

From~\eqref{P12},~\eqref{zw-def} it follows, that the two-point Green's function is determined by series expansions of the 4-point and 6-point correlation functions, $H_2$ and $H_4$, as the points $z_1,z_2\in\mathbb H$ and $z_3,z_4\in\mathbb H$ collapse pairwise. The series expansion of $H_2$ was determined in the previous section (see Eq.~\eqref{H2-expansion}). Therefore, we focus our attention on the 6-point correlation function,
\begin{equation}
H_4(z_1,z_2,z_3,z_4;x_1,x_2)= \langle \prod_{i=1}^4\Phi_{2,1}(z_i,\bar z_i) \Phi_{1,2}(x_1)\Phi_{1,2}(x_2)\rangle_{\mathbb H}.
\end{equation}
The leading order terms of the small-$\epsilon,\delta$ expansion of the correlation function are specified by possible channels of the fusion $\mathcal M_{2,1}\times\mathcal M_{2,1}$. As discussed, in the case of the bulk-bulk fusion we can use the OPE~\eqref{OPE-21} in order to obtain the series expansion of the correlation function:
\begin{multline}\label{H4-expansion}
H_4(z_1,z_2,z_3,z_4;x_1,x_2) = H_0(x_1,x_2)+ \\
+ \epsilon^{2/3}(C_{3,1})^2\langle\Phi_{3,1}(z,\bar z) \Phi_{1,2}(x_1)\Phi_{1,2}(x_2)\rangle_{\mathbb H} + \\
+ \delta^{2/3}(C_{3,1})^2\langle \Phi_{3,1}(w,\bar w)\Phi_{1,2}(x_1)\Phi_{1,2}(x_2)\rangle_{\mathbb H}+\\
+\epsilon^{2/3}\delta^{2/3}(C_{3,1})^4\langle \Phi_{3,1}(z,\bar z)\Phi_{3,1}(w,\bar w)\Phi_{1,2}(x_1)\Phi_{1,2}(x_2)\rangle_{\mathbb H}+O(\epsilon)+O(\delta).
\end{multline}
Upon substituting this expansion in~\eqref{P12} and taking account of~\eqref{H2-expansion} we determine the two-point SLE${}_{8/3}$ Green's function~\eqref{zw-def},
\begin{equation}\label{G-twopoint}
G_{\mathbb H}(z,w;x_1,x_2) = \frac{\langle \Phi_{3,1}(z,\bar z)\Phi_{3,1}(w,\bar w)\Phi_{1,2}(x_1)\Phi_{1,2}(x_2)\rangle_{\mathbb H}}{H_0(x_1,x_2)}.
\end{equation}
Hence, we conclude that the two-point SLE${}_{8/3}$ function can be written in terms of the correlation function in $c=0$ boundary LCFT.

Note, that the probability of the SLE${}_{8/3}$ trace to pass via two points~\eqref{zw-def} is expected to possess the following property: it should reduce to the one-point function~\eqref{G1-lim-def} when the points $z$ and $w$ collapse to one. In terms of the Green's function this property can be written as follows:
\begin{equation}\label{2G-1}
\lim_{\epsilon\to0} \epsilon^{2/3} G_{\mathbb H}\left(z-\frac{\epsilon \nu}{2},z+\frac{\epsilon\nu}{2};x_1,x_2\right) = c G_{\mathbb H}(z;x_1,x_2),
\end{equation}
where $\epsilon\ll1$, $|\nu|=1$, and $c$ is a constant. In the next section we will discuss this property in greater detail.

We also note, that the result for the two-point Green's function~\eqref{G-twopoint} can be easily generalized to the case of $N$ marked points in $\mathbb H$. We suggest the following expression for the multi-point Green's function:
\begin{equation}
G_{\mathbb H}(\{z_i\}_{i=1}^N;x_1,x_2)=\frac{(-1)^N\langle\prod_{i=1}^N \Phi_{3,1}(z_i,\bar z_i)\Phi_{1,2}(x_1)\Phi_{1,2}(x_2)\rangle_{\mathbb H}}{H_0(x_1,x_2)}.
\end{equation}

%%%%%%%%%%%%%%%%%%%%%%%%%%%%%%%%%%
\subsection{Coulomb gas representation of the Green's function}
%%%%%%%%%%%%%%%%%%%%%%%%%%%%%%%%%%

We end this section by proposing a Coulomb gas representation for the two-point Green's function\footnote{It can be generalized to the case of $N$ points.} in a somewhat heuristic manner. The correlation function on the right-hand side of~\eqref{G-twopoint} can be written as a linear combination of the conformal blocks,
\begin{multline}\label{H2cb-def}
\mathcal H_2(z,z^*,w,w^*;x_1,x_2;\gamma_1,\gamma_2,\gamma_3,\gamma_4)= \\
= \langle V_{3,1}(z) V_{3,1}(z^*) V_{3,1}(w) V_{3,1}(w^*) V_{1,2}(x_1) V_{-1,-2}(x_2)(Q^-)^4\rangle.
\end{multline}
The conformal blocks depend on the contours, $\gamma_1,\gamma_2,\gamma_3$, and $\gamma_4$, which determine the screening charges~\eqref{Qpm}.

\begin{figure}[t]
\centering
\includegraphics[width=1\columnwidth]{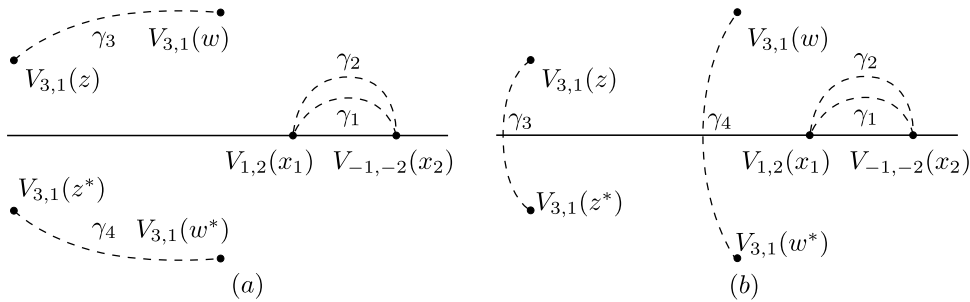}
\caption{\label{4_Green}
Two possible choices of the integration contours for the conformal block~\eqref{H2cb-def}. The dashed lines, $\gamma_i$, $i=1,2,3,4$, represent the integration contours, while the solid line denotes the boundary.
}
\end{figure}

Let us discuss possible choices of integration contours for~\eqref{H2cb-def}. Recall, that the contours are in one-to-one correspondence with the conformal blocks. The structure of $c=0$ boundary LCFT imposes strong constraints on the conformal blocks that contribute to the correlation function. In Ref.~\cite{SC09} it was argued that the theory must contain two logarithmic partners of the stress-energy tensor: $\Phi_{5,1}$ and $\Phi_{1,3}$. However, both fields (with different logarithmic couplings) can not appear in the theory simultaneously, because some quantities, e.g., $\langle\Phi_{1,3}\Phi_{5,1}\rangle$, are undefinable. Simmons and Cardy suggested that both fields can coexist provided that $\Phi_{5,1}$ appears in the bulk, while $\Phi_{1,3}$ --- on the boundary only. This conclusion imposes strong constraints on the bulk-boundary fusion. Namely, the bulk operators fuse to the boundary through the identity and the stress-energy tensor only.

This result suggests the following choice of the integration contours. Two contours, $\gamma_1$ and $\gamma_2$, should connect the operators $V_{1,2}(x_1)$ and $V_{-1,-2}(x_2)$ (see Fig.~\ref{4_Green}). By fusing these operators as $x_2\to x_1$, and shrinking the integration contours to a point in this process we obtain the screening charge $V_\alpha(x_1)$ with $\alpha=2\alpha_0+2\alpha_-$, so that $h_{2\alpha_0+2\alpha_-}=2$. This is the conformal dimension of the stress-energy tensor. Note, that this fusion agrees with previous results. Indeed, recall the conformal block~\eqref{H1-def} representing the one-point Green's function. Since the integration contour connects $z$ and $z^*$, the fusion of the vertex operators, $V_{1,2}(x_1)V_{1,2}(x_2)$ as $x_1\to x_2$, results in the operator $V_{\alpha}$ with $\alpha=2 \alpha_{1,2}$. Its conformal dimension, $h_{2\alpha_{1,2}}=2$, also equals the conformal dimension of the stress-energy tensor (see also Fig.~\ref{c-blocks}$(b)$).

Let us consider the other contours, $\gamma_3$ and $\gamma_4$, in the conformal block~\eqref{H2cb-def}. By requiring these contours to be symmetric with respect to the points $z,z^*,w$, and $w^*$, we consider two possibilities shown in Fig.~\ref{4_Green}: $(a)$ the contours connect the points $(z,w)$, and $(z^*,w^*)$, and ($b$) the contours connect $(z,z^*)$, and $(w,w^*)$. As discussed, the contours determine possible fusion channels, which contribute to the OPE of the field $\Phi_{3,1}$ and $\Phi_{3,1}$. Therefore, it is instructive to recall the fusion of the module $\mathcal M_{3,1}$ with itself. It reads~\cite{MR07}
\begin{equation}\label{M31M31}
\mathcal M_{3,1}\times \mathcal M_{3,1} = \mathcal M_{3,1} + \mathcal I_{5,1},
\end{equation}
where $\mathcal I_{5,1}$ is a staggered module, structurally described by the exact sequence $0\to\mathcal M_{1,1}\to\mathcal I_{5,1}\to\mathcal M_{5,1}\to0$. Note, that $\mathcal I_{5,1}$ is not itself a highest weight module. It is generated by the state $|\Phi_{5,1}\rangle$ with $h_{5,1}=2$, and the field $\Phi_{5,1}$ is a Jordan partner of the stress-energy tensor, $L_0|\Phi_{5,1}\rangle=2|\Phi_{5,1}\rangle+L_{-2}|0\rangle$, and $L_{2}|\Phi_{5,1}\rangle=-(5/8)|0\rangle$. Remarkably, the staggered module structure leads to the apeearance of logarithms in the correlation functions, e.g., $\langle\Phi_{5,1}(z)\Phi_{5,1}(0)\rangle=(5/4)\log(z)/z^4$.

Now, by taking account of the fusion rules~\eqref{M31M31} we discuss two blocks which can contribute to the correlation function~\eqref{H2cb-def} (see Fig.~\ref{4_Green}). In the case ($b$) the bulk-boundary fusion, $V_{3,1}(z)V_{3,1}(z^*)$ as $z,z^*\to x\in\mathbb R$, results in the vertex operator $V_\alpha(x)$ with the conformal dimension $h_{2\alpha_{3,1}+\alpha_-}=1/3$. It represents the boundary field $\Phi_{3,1}(x)$. However, we already noted at the begging of this section, that the bulk operators fuse to the boundary through the identity and the stress-energy tensor only. Therefore, the conformal block shown in Fig.~\ref{4_Green}($b$) is forbidden. In the case ($a$) the bulk-bulk fusion $V_{3,1}(z)V_{3,1}(w)$ as $w\to z$ results in the operator $V_{2\alpha_{3,1}+\alpha_-}(z)$ with $h_{2\alpha_{3,1}+\alpha_-}=1/3$. It corresponds to the bulk field, $\Phi_{3,1}(z)$, which generates the module $\mathcal M_{3,1}$ on the right-hand side of~\eqref{M31M31}. Besides, this fusion channel agrees with the limiting property of the two-point Green's function~\eqref{H2cb-def}.

The outlined reasoning suggests the following Coulomb gas representation of the two-point Green's function: it is determined by the conformal block shown in Fig.~\ref{4_Green}($a$). By computing the correlation function of vertex operators, we arrive with the following expression for the two-point Green's function:
\begin{multline}
G_{\mathbb H}(z,w;x_1,x_2)=\frac{X_2\eta_1^{2/3}(\eta_3-\eta_2)^{2/3}}{(z_1-\bar z_1)^{2/3}(z_2-\bar z_2)^{2/3}}\times\\
\times
\prod_{i=1}^3\frac{\eta_i^{4/3}}{1-\eta_i}\prod_{j<i}(\eta_i-\eta_j)^{4/3}\mathcal I_3(\eta_1,\eta_2,\eta_3).
\end{multline}
where $X_2$ is the normalization constant, $\eta_i=\eta(s_i)$ with $i=1,2,3$ are the cross-ratios~\eqref{eta-def} of the points $s=\{\bar z_1,z_2,\bar z_2\}$, and $\mathcal I_3$ denotes the 4-fold integral:
\begin{multline}
\mathcal I_3(\eta_1,\eta_2,\eta_3)=\int_{1}^{\infty}du_1 \int_{1}^{\infty}du_2 \int_{0}^{\eta_1}du_3 \int_{\eta_1}^{\eta_2}du_4\times \\
\prod_{i<j}(u_i-u_j)^{4/3}\prod_{i=1}^4 (u_i-1) u_i ^{-2/3}\prod_{j=1}^3(u_i-\eta_j)^{-2/3}.
\end{multline}
Here, the integration contours are obtained from the contours shown in Fig.~\ref{4_Green} by using the M\"obius transformation~\eqref{eta-def}.

%%%%%%%%%%%%%%%%%%%%%%%%%%%%%%%%%%%

\section{Conclusion and discussion}
In conclusion, let us briefly summarize the main results of the work. We considered the loop representation of the $O(n)$ model, and studied multi-point correlation functions of the twist operators, $\Phi_{2,1}$, in the bulk, and two 1-leg operators, $\Phi_{1,2}$, on the boundary on the upper-half plane. We used Dotsenko-Fateev method to obtain Coulomb gas representation of the $N\geq1$ correlation functions of the twist and legs operators. The correlation functions are written explicitly in terms of multi-fold contour integrals.

Afterwards, we focused upon the case $n=0$ representing self-avoiding loops in the $O(n)$ model. By following the Cardy-Simmons construction we connected the multi-point correlation functions with the probabilities of the SLE${}_{8/3}$ trace to wind in various ways about $N\geq 1$ points in the upper-half plane. Further, we obtained explicit expressions for the multi-point passage probabilities as linear combinations of the Coulomb gas integrals.

Afterward, we proposed a straightforward method to calculating the multi-point Green's functions of the SLE${}_{8/3}$ trace. By collapsing $2N$ marked points pairwise we recast the multi-point passage probabilities for the SLE trace in the $N$-point Green's functions. We showed that the Green's functions can be written in terms of the correlation functions in $c=0$ boundary LFCT containing the bulk operators $\Phi_{3,1}$, and a pair of the boundary 1-leg operators, $\Phi_{1,2}$. In the simplest case our construction leads to the well-known result for the one-point Green's function. By using heuristic arguments, we propose an explicit representation for the two-point Green's function. We are planning to elaborate this result in future publications.

%%%%%%%%%%%%%%%%%%%%%%%%%%%%%%%%%%%%

\section*{Acknowledgment}
The work is supported by the Russian Science Foundation grant 
19-71-30002.

%%%%%%%%%%%%%%%%%%%%%%%%%%%%%%%%%%%
\bibliographystyle{ieeetr}
\bibliography{biblio}{}
\end{document}